\documentstyle[12pt,ptp,epsf]{article}
\newfont{\bg}{cmr10 scaled\magstep4}
\newcommand{\bigzerol}{\smash{\hbox{\bg 0}}}
\newcommand{\bigzerou}{\smash{\lower1.ex\hbox{\bg 0}}}
\newcommand{\matrxt}[9]{\left(\begin{array}{ccc} #1 & #2 & #3 \\
                        #4 & #5 & #6 \\ #7 & #8 & #9 \end{array}\right)}
\newcommand{\matrxd}[4]{\left(\begin{array}{cc}  #1 & #2 \\
                        #3 & #4 \end{array}\right)}
\begin{document}
\preprint{
KANAZAWA 94-25,\\  
November, 1994  
}

\title{
Deconfinement transition and monopoles\\ in ${\bf T\neq0}$ ${\bf SU(2)}$ QCD
}
\author{
  Shun-ichi Kitahara$^{a}$ 
\footnote{ E-mail address:kitahara@hep.s.kanazawa-u.ac.jp}
, Yoshimi Matsubara$^{b}$
\footnote{ E-mail address:matubara@hep.s.kanazawa-u.ac.jp}
 and Tsuneo Suzuki$^{a}$
\footnote{ E-mail address:suzuki@hep.s.kanazawa-u.ac.jp}
}
\address{$^{a}$
Department of Physics, Kanazawa University, Kanazawa 920-11, Japan
}
\address{$^{b}$
Nanao Junior College, Nanao, Ishikawa 926, Japan
}

\maketitle

\begin{abstract}
The role of monopoles in the deconfinement transition 
is discussed in the framework of abelian projection in the 
maximally abelian gauge in  $T\neq0$ $SU(2)$ QCD. 
Only one (or a few near $\beta_c$) long connected monopole loop 
exists uniformly through the whole lattice in each vacuum 
configuration in addition to some very short loops in the confinement 
phase and the long loop disappears in the deep deconfinement region. 
Energy-entropy balance of the long loops of maximally extended monopoles 
explains the 
existence of the deconfinement transition and reproduces 
roughly the value of the critical temperature.
\end{abstract}

\vspace{1cm}

%\begin{center}
\section{Introduction}
%\end{center}

Modern computers enable us to simulate the system of quarks and gluons.
It has been evident from Monte-Carlo simulations of lattice QCD 
that quarks are confined.
However why and how quarks are confined is not yet known.
To understand the mechanism of the confinement 
is important in order to explain hadron physics out of QCD.
The 'tHooft idea of abelian projection of QCD is interesting.\cite{thooft}
The abelian projection is to fix the gauge in such a way that the maximal 
torus group remains unbroken.
After the abelian projection, monopoles appear as a topological quantity 
in the residual abelian channel.
QCD can be regarded as an abelian theory with electric charges and monopoles.
If the monopoles make Bose condensation, charged quarks and gluons 
are confined due to 
the dual Meissner effect.

There are, however, infinite ways of extracting such an 
abelian theory out of $SU(3)$ QCD. It seems important to find a 
good gauge in order to test whether the conjecture 
is true or not at least on a 
practically accessible lattice.
It has been found that, if one adopts a gauge called a maximally
abelian (MA) gauge,\cite{kron,yotsu,suzu93} the 't Hooft conjecture is 
seen to be beautifully realized in QCD.
\begin{enumerate}
\item 
The first interesting findings are some phenomena called abelian dominance.
Abelian loop operators composed of abelian link variables alone
 seem to reproduce essential features of color confinement
in the MA gauge.\cite{yotsu,hio} 
Explicitly, abelian static potentials derived from 
abelian Wilson loops can reproduce 
the string tensions which is a key quantity of confinement.
Abelian Polyakov loops and abelian 
energy densities play the role of an order parameter 
of the deconfinement transition in finite-temperature 
pure QCD.\cite{suzu93}
The abelian quantities show clearer behaviors
around the critical coupling. 
\item 
Furthermore, it has been shown that monopoles alone 
are responsible for the above interesting phenomena played 
by the abelian quantities. Abelian Wilson 
loops\cite{shiba2,shiba3,shiba4,ejiri2,ejiri3}
 and abelian Polyakov loops\cite{suzu94a,matsu} are written by a product of 
monopole and photon contributions. The string tension and the behavior of 
the Polyakov loop as an order parameter of the deconfinement transition 
are all reproduced by the monopole contributions alone.
\item
Abelian dominance suggests that there must exist a $U(1)$ invariant 
effective action which can explain confinement. It is possible to get the 
$U(1)$ effective action in terms of monopole currents in 
$T=0$ $SU(2)$ QCD after performing 
a dual transformation numerically.\cite{shiba1,shiba3,shiba5,shiba6,suzu94b}
The effective action is fixed also for extended 
monopoles.\cite{ivanenko} Considering extended monopoles corresponds 
to making a block spin transformation on the dual lattice. The actions 
determined are very interesting because they appear to satisfy 
a scaling behavior
of the renormalized trajectory on which one can take the continuum limit.
\cite{shiba1,shiba3,shiba6,suzu94b}   
If the similar behaviors would show up on a larger lattice, one could 
conclude that $T=0$ $SU(2)$ QCD is always (
for all $\beta$ in the infinite lattice limit) 
 in the monopole condensed phase. Color confinement would be proved then.
\end{enumerate}

What happens in the case of the deconfinement transition in the 
finite-temperature ($T\neq 0$) QCD?  Can the similar 
monopole dynamics like the energy-entropy balance explain the transition?
This is the main subject of this note. It is highly probable, since 
the monopole contributions alone explain the string tension and the behavior 
of the Polyakov loop also in $T\neq 0$ QCD.\cite{ejiri2,ejiri3,suzu94a,matsu}

In the next section, what is abelian projection is shortly reviewd. 
Section 3 reviews the results in $T=0$ QCD. Section 4 is the most important 
part of this note discussing the case of $T\neq0$ $SU(2)$ QCD.
Final section is devoted to conclusion and remarks.

%\begin{center}
\section{Abelian projection}
%\end{center}
\subsection{Abelian projection in the continuum QCD}
Abelian projection of QCD is done as follows.
Choose an operator $X(x)$ which is transformed non-trivially 
under $SU(3)$ transformation:
\begin{eqnarray}
   A_\mu (x) & \to & \widetilde A_\mu (x)= V(x)A_\mu (x)V^\dagger (x)
   - \frac{i}{g}\partial_\mu V(x)V^\dagger (x)\\
   \psi (x)& \to & \widetilde \psi (x)= V(x)\psi (x).
\end{eqnarray}
Abelian projection is to choose $V(x)$ so that $X(x)$ is diagonalized as 
\begin{eqnarray}
   X(x) \to \widetilde X(x)
   = \matrxt{\lambda_1 (x)}{}{\bigzerou}{}
            {\lambda_2 (x)}{}{\bigzerol}{}{\lambda_3 (x)} .
\end{eqnarray}
$V(x)$ is fixed up to the diagonal element $d(x)$ of $SU(3)$, where
\begin{eqnarray}
   d(x) = \matrxt{\exp(i\alpha_1 (x))}{}{\bigzerou}{}
              {\exp(i\alpha_2 (x))}{}{\bigzerol}{}{\exp(i\alpha_3 (x))}
     \quad \in SU(3) , \qquad \sum_{i=1}^{3}\alpha_i (x)=0.
\end{eqnarray}
$\{d(x)\}$ is the maximum torus group of $SU(3)$, which is the residual 
$U(1)\times U(1)$ gauge symmetry.

We look at QCD at this stage without further fixing the gauge 
of the residual symmetry. 
First, we explore how the fields after the abelian projection 
transform 
{\it under an arbitrary $SU(3)$ gauge transformation $W(x)$.}
 Since 
$X(x)$ is a functional of (gauge) fields and so it 
transforms under $W(x)$. Hence $V(x)$ also transforms non-trivially.
Let us fix the form of $V(x)$ such that all diagonal components 
of the exponent of $V(x)$ are zero. This is always possible if one uses the 
residual symmetry. Then 
$V(x)$ is found to transform 
{\it under $W(x)$}
 as
\begin{eqnarray}
   V(x)\stackrel{W}{\longrightarrow}V^W (x)= d^W (x)V(x)W^\dagger (x).
\end{eqnarray}
$V^W (x)$ diagonalizes an operator which is transformed from 
$X(x)$ under $W(x)$.
$d^W (x)$ is necessary for $V^W (x)$ to take the fixed form.

The gauge field after the abelian projection, 
$\widetilde A_\mu (x)$, transforms 
{\it under $W(x)$} as
\begin{eqnarray}
   \widetilde A_\mu (x)\stackrel{W}{\longrightarrow}\widetilde A_\mu^W (x) = 
d^W (x)\widetilde A_\mu (x)d^{W\dagger}(x)
   - \frac{i}{g}\partial_\mu d^W (x)d^{W\dagger}(x). \label{atran}
\end{eqnarray}
After the abelian projection, $\widetilde A_\mu (x)$  
transforms only under the diagonal matrix $d^W (x)$.
Since the last term of (\ref{atran}) is composed of the diagonal part alone, 
the diagonal part of $\widetilde A_\mu (x)$ transforms like a photon.
The off diagonal part of $\widetilde A_\mu (x)$ transforms 
like a charged matter.
The quark field transforms {\it under $W(x)$} as
\begin{eqnarray}
   \widetilde\psi (x)\stackrel{W}{\longrightarrow}\widetilde\psi^W (x)
= d^W (x)\widetilde\psi (x).
\end{eqnarray}
It is important that $\widetilde\psi^\dagger_i (x)\widetilde\psi_i (x)$ and 
$\widetilde\psi_1 (x)\widetilde\psi_2 (x)\widetilde\psi_3 (x)$ 
are neutral and 
at the same time  invariant under any $SU(3)$ transformation $W(x)$.
{\it Color confinement is regarded as abelian charge confinement after abelian 
projection.}

The most interesting fact of abelian projection is that monopoles appear 
in the residual abelian channel.
We treat $SU(2)$ QCD for simplicity. 
After abelian projection, we define an abelian field strength as 
\begin{eqnarray}
   f_{\mu\nu}(x) = \partial_\mu \widetilde A_\nu^3 (x) 
              - \partial_\nu \widetilde A_\mu^3 (x).
\end{eqnarray}
The abelian field $\widetilde A_\mu^3 (x)$ written in terms of the original 
field is
\begin{eqnarray}
   \widetilde A_\mu^3 (x)= \hat Y^a (x)A_\mu^a (x)
     - \frac{1}{g}\frac{1}{1+\hat Y^3 (x)}\varepsilon_{3ab}
       \hat Y^a (x)\partial_\mu \hat Y^b (x),
\end{eqnarray}
where $\hat Y(x) =V^{\dagger}(x) \sigma_3 V(x)=\hat Y^a (x)\sigma^a$.
$\hat Y^a (x)$ obeys 
\begin{equation}
\hat Y^a (x)\hat Y^a (x)=1.\label{ysphere}
\end{equation}
$f_{\mu\nu}(x)$ can be rewritten in the form
\begin{eqnarray}
   f_{\mu\nu} (x)= \partial_\mu (\hat Y^a (x)A_\nu^a (x))
-\partial_\nu (\hat Y^a (x)A_\mu^a (x))
              - \frac{1}{g}\varepsilon_{abc}\hat Y^a (x)
\partial_\mu \hat Y^b (x)\partial_\nu \hat Y^c (x)
\end{eqnarray}
in terms of the original field.
A current
\begin{eqnarray}
   k_{\mu}(x) &=& \frac{1}{2}\varepsilon_{\mu\nu\rho\sigma}\partial^\nu
             f^{\rho\sigma}(x)         \\
         &=& \frac{1}{2g}\varepsilon_{\mu\nu\rho\sigma}
             \varepsilon_{abc}\partial^\nu\hat Y^a (x)
\partial^\rho \hat Y^b (x)\partial^\sigma \hat Y^c (x)
\end{eqnarray}
is always zero if $V(x)$ is fixed. However,
at a point $x$ where the eigenvalue of the diagonalized operator 
$X(x)$ is degenerate, $V(x)$ is not well defined and $k_{\mu}(x)$ does 
not vanish there.
We calculate the charge in the three dimensional volume 
$\Omega$ around $x$:\cite{arafune}
\begin{eqnarray}
   g_m=\int_\Omega k_0 (x)d^3x
     &=& \frac{1}{2g}\int_\Omega\varepsilon_{0\nu\rho\sigma}
         \varepsilon_{abc}\partial^\nu\hat Y^a (x)\partial^\rho \hat Y^b (x)
         \partial^\sigma \hat Y^c (x)d^3x \\
     &=& \frac{1}{2g}\int_{\partial\Omega}\varepsilon_{ijk}
         \varepsilon_{abc}\hat Y^a (x)\partial_j \hat Y^b (x)
         \partial_k \hat Y^c (x)d^2\sigma_i \\
     &=& \frac{4\pi n}{g}  ,
\end{eqnarray}
where $n$ is an integer.
$n$ is a topological number corresponding to a mapping between 
the sphere (\ref{ysphere}) in the parameter space and the sphere 
$\partial\Omega$ of $\Omega$.
Because this equation represents the Dirac quantization condition, 
$g_m$ can be interpreted as a magnetic charge.
The monopole current $k_\mu (x)$ is a topologically conserved current
  $ \partial_\mu k^\mu (x) = 0 .$
{\it Abelian projected QCD can be regarded as 
an abelian theory with electric charges and monopoles.}
'tHooft\cite{thooft} conjectured
if the monopoles condense, abelian charges are confined 
due to the dual Meissner effect. This means color confinement.

\subsection{Abelian projection on a lattice}
We can perform abelian projection on a lattice similarly.
Choose an operator $X(s)$ in, for simplicity, $SU(2)$ QCD.
The gauge transformation on a lattice is  
\begin{eqnarray}
   \widetilde U(s,\hat\mu)=V(s)U(s,\hat\mu)V^\dagger (s+\hat\mu),
\end{eqnarray}
where $U(s,\hat\mu)$ represents a link field corresponding to a gauge field 
in the continuum theory.
After abelian projection is over,  abelian link fields can be separated 
from  $SU(2)$ link fields as follows:
\begin{eqnarray}
   \widetilde U(s,\hat\mu)
         &=&\matrxd{\sqrt{1-\left| c(s,\mu)\right|^2}}{-c^\ast (s,\mu)}
                 {c(s,\mu)}{\sqrt{1-\left| c(s,\mu)\right|^2}}
            \matrxd{e^{i\theta_\mu (s)}}{0}{0}{e^{-i\theta_\mu (s)}}  \\
         &=&C(s,\hat\mu)u(s,\hat\mu)  .
\end{eqnarray}
$u(s,\hat\mu)$ transforms like a photon and $C(s,\hat\mu)$ transforms 
like a charged matter under the residual $U(1)$ gauge symmetry.
The abelian field strength is defined as a plaquette variable 
\begin{eqnarray}
   \theta_{\mu\nu}(s)=\theta_\mu(s)+\theta_\nu(s+\hat\mu)
                     -\theta_\mu(s+\hat\nu)-\theta_\nu(s).
\end{eqnarray}
The monopole current is defined\cite{degrand} as
\begin{eqnarray}
   k_\mu(s)=\frac{1}{2}\varepsilon_{\mu\nu\rho\sigma}\partial_\nu
            n_{\rho\sigma}(s+\hat\mu) ,
\label{eqn:latk}
\end{eqnarray}
where $\partial_\nu$ is a forward derivative on a lattice and 
$\theta_{\mu\nu}(s)$ is decomposed into 
\begin{eqnarray}
   \theta_{\mu\nu}(s)=\bar\theta_{\mu\nu}(s)+2\pi n_{\mu\nu}(s), \quad 
   -\pi<\bar\theta_{\mu\nu}(s)\leq\pi.
\end{eqnarray}
$n_{\mu\nu}(s)$ is an integer corresponding to the number of the Dirac string 
through the plaquette.
The monopole currents are conserved topologically
\begin{eqnarray}
   \partial_\mu 'k_\mu = 0 ,\label{cons}
\end{eqnarray}
where $\partial_\mu '$ is a backward derivative on a lattice.
The monopole currents make closed loops on the four dimensional lattice.
%----------- Fig.1----------------
\begin{figure}[tbh]
%\begin{minipage}[t]{7cm}
\epsfxsize=.4\textwidth
\begin{center}
\leavevmode
\epsfbox{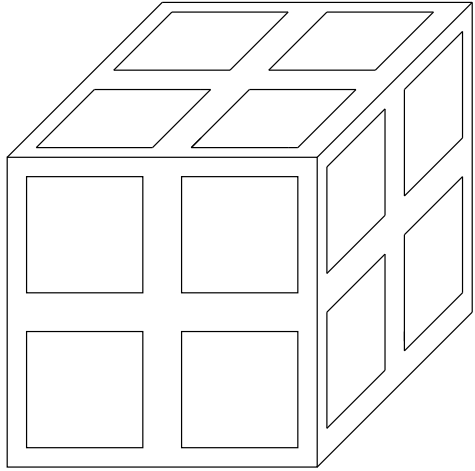}
\end{center}
\caption{
An extended cube on which an $2^3$ extended monopole is defined as the sum of 
eight (=$2^3$) smallest monopoles. 
}
\label{f1}
%\end{minipage} \   \ 
\vspace{1cm}
\end{figure}
%---------------------------------
The above monopole current is defined by 
$n_{\mu\nu}(s)$ surrounding 
the smallest cube as shown in (\ref{eqn:latk}).
To study the long range behavior important in QCD, considering 
extended monopoles\cite{ivanenko} is 
essential.\cite{shiba1,shiba3,shiba6,suzu94b} They are 
defined by 
$n_{\mu\nu}(s)$ surrounding an extended cube.
For example, Fig.\ref{f1} represents a $2^3$ cube defining a
$2^3$ extended monopole.
Adopting a $n^3$ extended monopole corresponds to performing a block spin 
transformation on a dual lattice\cite{shiba1,shiba6,suzu94b} and so is 
suitable for exploring the long range property of QCD.
When adopting a $n^3$ extended monopole on $N_s^3\times N_t$ lattice, 
the effective lattice on which the extended monopole runs is 
\begin{eqnarray}
   \left( \frac{N_s}{n}\right)^3\times\left( \frac{N_t}{n}\right) 
\end{eqnarray}
which we call a renormalized lattice.

%\begin{center}
\section{Monopole condensation in $T=0$ QCD}
%\end{center}
As shown in the introduction, a gauge called a maximally
abelian (MA) gauge\cite{kron,yotsu,suzu93} is 
very interesting. In the MA gauge,
\begin{eqnarray}
   X(s)=\sum_\mu \left[ U(s,\hat\mu )\sigma_3U^\dagger (s,\hat\mu )
        +U^\dagger (s-\hat\mu,\hat\mu )\sigma_3U(s-\hat\mu,\hat\mu )
        \right]
\end{eqnarray}
is diagonalized. 
Abelian loop operators composed of $u(s,\hat\mu )$ alone 
seem to reproduce essential features of color confinement.
Here we review briefly the results 
showing that monopoles are a key quantity of confinement in $T=0$ $SU(2)$
QCD. They are obtained by Shiba and 
one of the authors (TS) 
recently.\cite{shiba1,shiba2,shiba3,shiba4,shiba5,shiba6,suzu94b} 

The first interesting result is that one can determine an effective 
$U(1)$ action in terms of monopole currents in the MA gauge 
in $SU(2)$ QCD.\cite{shiba1,shiba3,shiba5,shiba6}
The partition function of interacting monopole currents is expressed as
\begin{eqnarray}
  Z=(\prod_{s,\mu} \sum_{k_{\mu} (s)=-\infty}^\infty ) \,
    (\prod_{s} \delta_{\partial '_{\mu}k_{\mu}(s) ,0 } ) \,
    \exp (-S[k]) .
\label{eqn:pfunm}
\end{eqnarray}
It is natural to assume $S[k] = \sum_i f_i S_i [k]$. Here
$f_i$ is a coupling constant of an interaction $S_i [k]$. 
For example, $f_1$ is the coupling of the self energy term  
$\sum_{n,\mu}(k_\mu(s))^2$, $f_2$ is the coupling 
of a nearest-neighbor interaction term
 $\sum_{n,\mu} k_\mu(s) k_\mu(s+\hat\mu)$ 
and $f_3$ is the coupling of another nearest-neighbor term 
$\sum_{n,\mu\neq\nu} k_\mu(s) k_\mu(s+\hat\nu)$.\cite{shiba1,shiba6}
Shiba and Suzuki\cite{shiba1,shiba3,shiba5,shiba6} extended a method
developed by Swendsen\cite{swendsn} to the system of monopole currents 
obeying the current conservation rule (\ref{cons}).
The monopole actions are obtained locally enough 
for all extended monopoles considered even in the scaling region. They are 
lattice volume independent.
The coupling constant $f_1$ of the self-energy term is dominant 
and the coupling constants decrease 
rapidly as the distance between the two monopole currents increases.

%----------- Fig.2----------------
\begin{figure}
\vspace{-1cm}
\epsfxsize=0.6\textwidth
\begin{center}
\leavevmode
\epsfbox{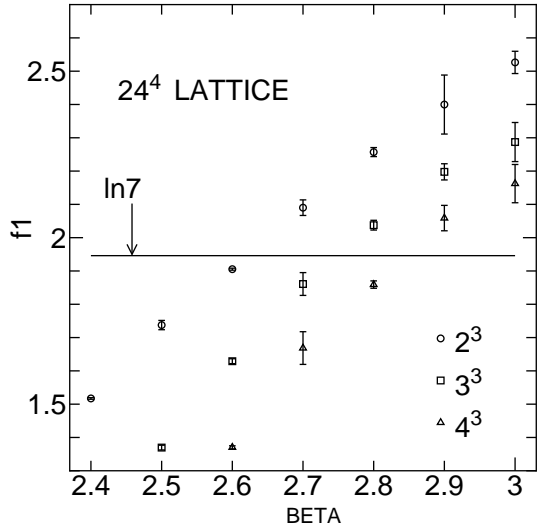}
\end{center}
\vspace{-1cm}
\caption{
Coupling constants $f_1$ versus $\beta$ for 
$2^3,3^3$, and $4^3$ 
extended monopoles on $24^4$ lattice. 
}
\label{f2}
\end{figure}
%---------------------------------

Since the action is fixed, it is possible to study 
energy and entropy balance of monopole loops in order to 
confirm the occurence of  monopole condensation.
If the entropy of a monopole loop exceeds the energy, 
the condensation of a monopole loop  occurs.
As done in compact QED,\cite{bank} 
the entropy of a monopole loop can be estimated as $\ln 7$ 
per unit loop length.
Since monopole currents are distributed randomly in average 
for large $L$, interaction terms between two separate  
currents is expected to be canceled. That this actually happens will be 
shown later in the case of $T\neq 0$ QCD case. Hence 
the action may be approximated by 
the self energy part $f_1 L$. Here 
dominance of currents with a unit charge is used.
Since $f_1$ is regarded as the self energy per unit monopole loop length, 
the free energy per unit monopole loop length is approximated by 
\begin{eqnarray}
   f_1-\ln 7  .
\end{eqnarray}
If $f_1 < \ln 7$, the entropy dominates over the energy, 
which means condensation of monopoles.
In Fig.\ref{f2}, $f_1$ versus $\beta$ for various extended 
monopoles on $24^4$ 
lattice is shown in comparison with the entropy value $\ln 7$.
Each extended monopole has its own $\beta$ region where the condition
$f_1 < \ln 7$ is satisfied.
When the extendedness is bigger, larger $\beta$ is included in such a 
region. Larger extended monopoles are more important in determining 
the phase transition point.

%----------- Fig.3----------------
\begin{figure}
\vspace{-1cm}
\epsfxsize=0.6\textwidth
\begin{center}
\leavevmode
\epsfbox{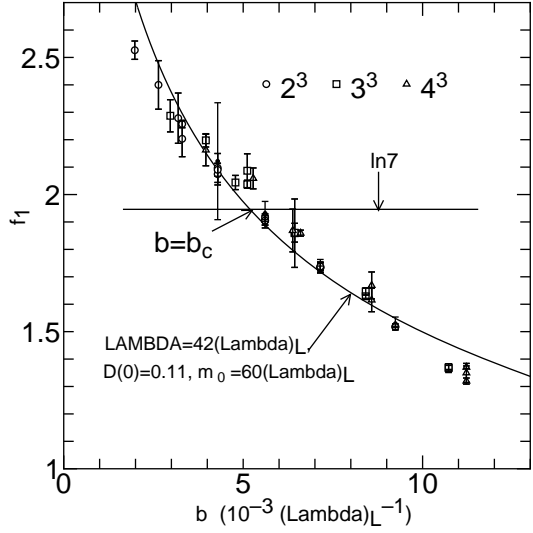}
\end{center}
\vspace{-1cm}
\caption{
Coupling constants $f_1$ versus $b$.
}
\label{f3}
\end{figure}
%---------------------------------

The behaviors of the coupling constants are 
different for different extended 
mono-poles. However, 
if we plot them versus 
$b=n\times a(\beta)$,
we get a unique curve as in Fig.\ref{f3}. The 
coupling constants seem to depend only on $b$, not on the extendedness 
nor $\beta$.  
There is a critical $b_c$ corresponding to critical $\beta^n_c$,
 i.e., $b_c =na(\beta^n_c)$.
The monopole action may be fitted by
\begin{eqnarray}
S[k] & = & \sum m_0 b k_{\mu}(s)k_{\mu}(s) 
+ \frac{1}{2}(\frac{4\pi}{g(b)})^2 \sum 
k_{\mu}(s)\bar{D}(s-s')k_{\mu}(s'), 
\label{monoact}
\end{eqnarray}
where $g(b)$ is the SU(2) running coupling constant
\begin{eqnarray}
g(b)^{-2}= \frac{11}{24\pi^2}\ln(\frac{1}{b^2\Lambda^2}) 
+ \frac{17}{44\pi^2}
\ln\ln(\frac{1}{b^2\Lambda^2}).
\end{eqnarray}
$\bar{D}(s)$ is a 
modified lattice Coulomb propagator.
This form of the action is predicted 
theoretically by Smit and Sijs.\cite{smit}
The solid line is the prediction given 
by the action with the parameters 
written in Fig.\ref{f3}.

Suppose the effective monopole action remains the same for any 
extended
 mono-poles larger than $4^3$ in the infinite volume 
limit. Then the finiteness of $b_c =na(\beta^n_c)$ suggests 
$\beta^n_c$ becomes infinite when the extendedness $n$ 
goes to infinity. $SU(2)$ lattice QCD is always (for all $\beta$) 
in the monopole condensed and then in the color confinement phase.\cite{thooft} This is one of what one wants to prove in the framework of lattice QCD.

Notice again that considering extended monopoles correponds to performing 
a block spin transformation on the dual lattice. 
The above fact that the effective actions 
for all extended monopoles considered 
are the same for fixed $b$ means that the action may be  
the renormalized trajectory on which one can take the continuum limit.
Our results suggest the continuum monopole action takes the form
 (\ref{monoact}) predicted by Smit and Sijs.\cite{smit}
The simulation of the monopole action is in progress.

Shiba and Suzuki\cite{shiba2,shiba3,shiba4} showed furthermore 
that monopoles alone can reproduce the full 
value of the string tension in $SU(2)$ QCD.
These facts strongly suggest that the monopole condensation occurs 
in QCD and quarks and gluons are confined due to the dual Meissner effect.

%\begin{center}
\section{Deconfinement transition in $T\neq 0$ QCD}
%\end{center}

\begin{figure}
\epsfxsize=0.6\textwidth
\begin{center}
\leavevmode
\epsfbox{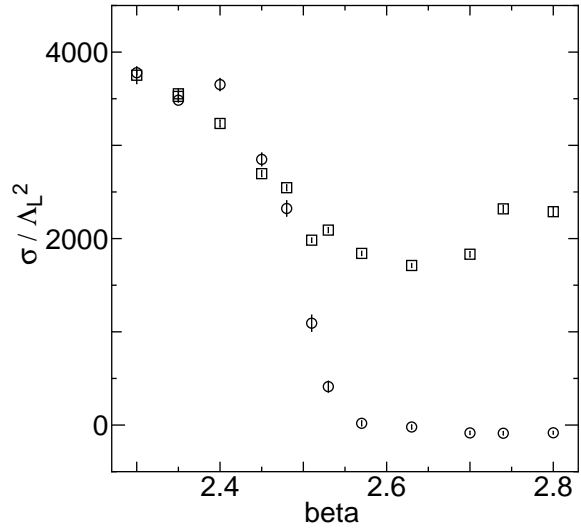}
\end{center}
\vspace{-1cm}
\caption{
Physical string tensions (circle) and spatial string tensions (square)
from monopoles on $24^{3} \times 8$ lattice.
}
\label{f4}
\end{figure}

Now let us study 
the role of the monopoles in the deconfinement transition 
in $T\neq 0$ QCD. 
There have been interesting data already 
suggesting the importance of monopoles also 
in $T\neq 0$ QCD. 
Monopole contribution to the string tension has been studied 
 in the MA gauge also in $T\neq 0$ $SU(2)$ QCD.
The string tension from the monopoles almost 
agrees with that from the abelian Wilson loops in the confinement phase, 
whereas it vanishes clearly in the deconfinement phase.
The data are shown in Fig.\ref{f4}.
The string tension from the photon part is negligibly small.

%----------- Fig.4----------------
%---------------------------------

%----------- Fig.5----------------
\begin{figure}
\vspace{-1cm}
\epsfxsize=0.6\textwidth
\begin{center}
\leavevmode
\epsfbox{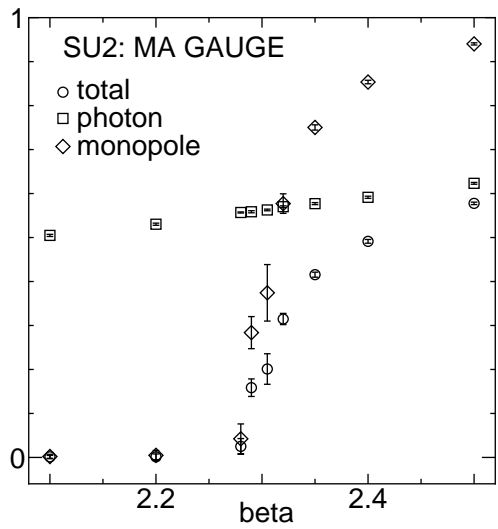}
\end{center}
\vspace{-1cm}
\caption{
Monopole Dirac string and photon contributions to Polyakov loops in the 
MA gauge in $SU(2)$ QCD on $16^3\times 4$ lattice. Total means the 
abelian Polyakov loops.
}
\label{f5}
\end{figure}
%---------------------------------

An abelian Polyakov loop which is written in terms of abelian link fields 
alone is given by a product of contributions from Dirac strings 
of monopoles and from photons. It has been found \cite{suzu94a,matsu}
that the characteristic features of the Polyakov loops as an order 
parameter of the deconfinement transition are 
due to the Dirac string contributions alone.
In Fig.\ref{f5}, the $SU(2)$ data in the MA gauge are plotted.
The abelian Polyakov loops vanish in the confinement phase 
whereas they begin to rise for $\beta$ larger than 
the critical temperature $\beta_c=$ 2.298.
The Dirac string contribution shows similar behaviors more drastically.
The photon part has a finite contribution for both phases and 
it changes only slightly.
The fact that monopoles are responsible for the essential feature of the 
Polyakov loop is found also in $U(1)$ and in $SU(3)$ in the MA gauge.
In addition, a remarkable result has been found.
The behavior of the abelian Polyakov loops as an order parameter and the 
monopole responsibility are seen also 
{\it in other gauges like unitary ones.}
This is the first phenomena suggesting gauge independence of 
the 'tHooft conjecture.

\subsection{The expected role of maximally extended 
monopoles 
}
In $T=0$ QCD, 
each extended monopole has its own $\beta$ region where 
the entropy dominance of monopoles occurs, i.e., where the condition
$f_1 < \ln 7$ is satisfied.
When the extendedness is bigger, larger $\beta$ is included in such a 
region. Larger extended monopoles are more important in determining 
the phase transition point. In the infinite-volume limit, 
one can adopt infinitely extended monopoles. If the situations remain 
unaltered even in the infinite-volume limit, one can prove that
$T=0$ $SU(2)$ lattice QCD is always (for all $\beta$) 
in the monopole condensed and then in the color 
confinement phase.\cite{thooft} 

What happens in $T\neq 0$ QCD? 
It is expected that the confinement - deconfinement transition 
in $T\neq 0$ QCD also can be explained in terms of monopoles.
Then energy and entropy balance of monopoles 
must be the mechanism as shown above in $T=0$ QCD.
However, there is a big difference between $T=0$ and $T\neq 0$ QCD. 
Namely the time extent $N_t$ is kept finite in $T\neq 0$ case.
This means that the extendedness of monopoles running in the space 
direction (which we call dynamical monopoles) is restricted to 
be finite, since monopoles are defined by a three dimensional 
cube perpendicular to the direction of the current. 
The monopole currents which contribute to the physical string tension are 
those running in the direction perpendicular to the Wilson loops.
Because the Wilson loops are on a plane which include the time axis 
and a space one, dynamical monopoles are  
essential in the confinement mechanism. It is our expectation that 
{\it 
the existence of the maximum size of extended dynamical monopoles 
is the reason there is a deconfinement phase transition at 
a finite value of 
 $\beta_c$, i.e., $T_c$ in the finite temperature QCD. 
}
Here $T_c$ is the critical temperature.

In the following subsections, we explore what is the maximum size of 
extended monopoles 
and whether $\beta_c$ can be explained by balancing of energy and 
entropy of the maximally extended monopole loops in the MA gauge.

%----------- Fig.6----------------
\begin{figure}
\vspace{-1cm}
\epsfxsize=0.6\textwidth
\begin{center}
\leavevmode
\epsfbox{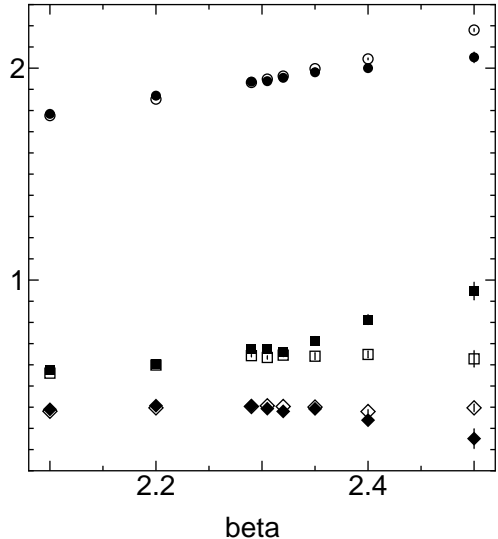}
\end{center}
\vspace{-1cm}
\caption{
Coupling constants $f_1$(circle), $f_2$(square) and $f_3$(diamond) 
versus $\beta$ on $16^3\times 4$ lattice for $1^3$ monopoles.
Open symbols are the couplings connecting two space-like currents 
and filled ones are those connecting two time-like currents.
}
\label{f6}
\end{figure}
%---------------------------------

\subsection{Monopole action in $T \neq 0$ QCD}
Monopole action can be calculated from 
monopole current configurations similarly as in $T=0$ $SU(2)$ QCD.
The partition function is expressed by Eq.(\ref{eqn:pfunm}).
In contrast with the case of $T=0$ QCD, we must treat 
time and space directions separately.
Since the number of sites in the time direction is small, 
we consider only up to next to the nearest quadratic couplings.
Owing to the current conservation, five interaction terms are 
sufficient in this case.
In Fig.\ref{f6} the action is calculated in both the confinement 
and the deconfinement phases on $16^3 \times 4$ lattices. We used  
50 (100 at some points) configurations adopting $1^3$ monopoles.
$f_1$ is the coupling of the self energy term 
$\sum_{n,\mu}(k_\mu(s))^2$, $f_2$ is the coupling 
of the term $\sum_{n,\mu} k_\mu(s) k_\mu(s+\hat\mu)$ 
and $f_3$ is the coupling of the term, 
$\sum_{n,\mu\neq\nu} k_\mu(s) k_\mu(s+\hat\nu)$.
Other couplings between currents larger distance apart
are much smaller.
$f_1$ is dominant in both phases.
In the deconfinement phase, the discrepancy between 
space-space and time-time couplings is large, 
whereas it is negligible in the confinement phase.

%----------- Fig.7----------------
\begin{figure}
\vspace{-1cm}
\epsfxsize=0.6\textwidth
\begin{center}
\leavevmode
\epsfbox{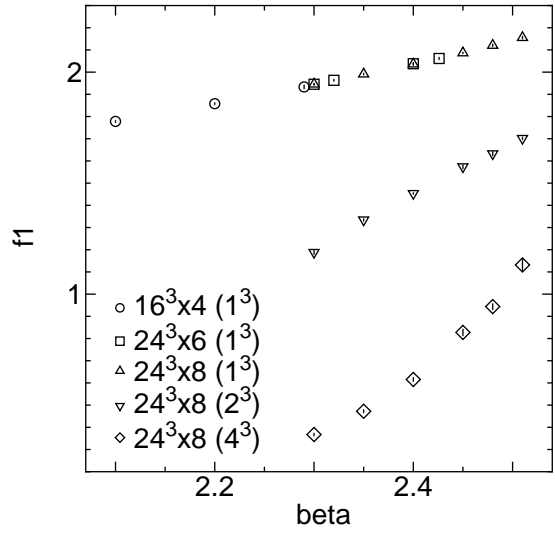}
\end{center}
\vspace{-1cm}
\caption{
Coupling constants $f_1$ in the confinement phase versus $\beta$ 
on various lattices.
The parenthesized number represents the extendedness of monopoles.
}
\label{f7}
\end{figure}
%---------------------------------

$f_1$ of the monopole action is plotted in Fig.\ref{f7} in the confinement 
phase on various lattices.
In the confinement phase the action is 
independent of $N_s$ and $N_t$ ,i.e., of the lattice volume  
and the action obtained seems the same as that given in $T=0$ QCD.

%----------- Fig.8----------------
\begin{figure}
\vspace{-1cm}
\epsfxsize=0.6\textwidth
\begin{center}
\leavevmode
\epsfbox{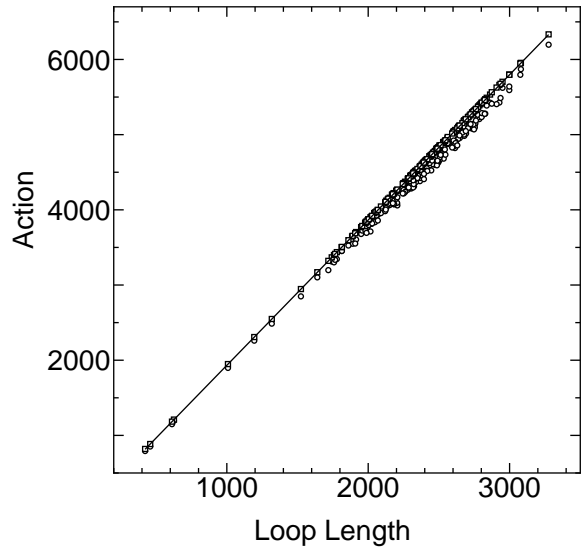}
\end{center}
\vspace{-1cm}
\caption{
The values of the total monopole action (circle), 
self energy term of the action (square) and $f_1$ times the length of 
a monopole loop $L$ versus monopole loop length (solid line).
}
\label{f8}
\end{figure}
%---------------------------------

Now that the energy of a monopole loop is evaluated, let us check if 
all terms other than 
the self energy are canceled because of randomness of 
the monopole loop distribution.
In Fig.\ref{f8} we compare the value of the total 
monopole action with that of 
the self energy term and also with that of $f_1 \times L$, 
where $L$ is the length of a monopole loop.
The length of monopole loops will be defined  
in detail in the next subsection.
The data show that the cancellation is good and the total action is 
well approximated only by the self energy term.
Furthermore the value of the self energy term is in fairly good agreement 
with 
$f_1\times L$, which suggests dominance of monopole currents with unit charge.
Next, we pay attention to the entropy of a monopole loop.

\subsection{Monopole loop length}
Suppose that the behavior of a monopole loop is represented by 
non-backtracking random walks, 
the number of loops of length $L$ on four-dimensional large lattice 
behaves as $7^L$.
 The entropy of a monopole loop can be estimated  
as $\ln 7$ per unit loop length in $T=0$ QCD.
However, the entropy of the loop in $T\neq 0$ QCD must be affected 
by the finiteness of the time direction and the entropy may become 
smaller than $\ln 7$.
To estimate the entropy, we have to investigate the behavior 
of monopole loops carefully 
in $T\neq 0$ QCD.

%-------------- table 1 ----------------
\begin{table}[tb]
\begin{center}
\begin{tabular}{cc|cc|cc|cc} 
(a) &        &     &        &     &        &     &        \\       
$L$ & number & $L$ & number & $L$ & number & $L$ & number \\ \hline
     4 & 708 &  24 &   4 &    46 &   1  &  5010 &   1  \\
     6 & 238 &  26 &   1 &    52 &   1  &  5012 &   1  \\
     8 &  89 &  28 &   1 &   100 &   1  &  5092 &   1  \\
    10 &  50 &  30 &   3 &  4612 &   1  &       &      \\
    12 &  33 &  32 &   4 &  4798 &   1  &       &      \\
    14 &  24 &  34 &   3 &  4822 &   1  &       &      \\
    16 &  15 &  36 &   1 &  4854 &   1  &       &      \\
    18 &   8 &  40 &   2 &  4876 &   1  &       &      \\
    20 &   9 &  42 &   1 &  4880 &   1  &       &      \\
    22 &   6 &  44 &   1 &  4940 &   1  &       &      \\ \hline\hline
(b) &        &     &        &     &        &     &        \\ 
$L$ & number & $L$ & number & $L$ & number & $L$ & number \\ \hline
    2 &   19 &   10 &    1 & 1292 &    1 & 1336 &    1 \\
    4 &   14 & 1272 &    2 & 1302 &    1 & 1340 &    2 \\
    6 &    2 & 1288 &    1 & 1318 &    1 & 1348 &    1 \\
\end{tabular}
\end{center}
\caption{
The total number of loops of length $L$ 
on a sample of 10 configurations on $16^3\times 4$ lattice at $\beta=$2.20
for $1^3$ monopoles (a) and $2^3$ monopoles (b).
}
\label{t1}
\end{table}
%-------------------------------------

First, we calculate the length of monopole loops.
Bode et al.\cite{bode} studied the length of monopole loops in $U(1)$ lattice 
gauge theory.
They found that in the confinement phase one large loop occupies 
the whole lattice and in the Coulomb phase it splits into 
several smaller pieces.
We define  the monopole loop length following Bode et al\cite{bode}
in the MA gauge in $SU(2)$ QCD.
If two closed loops occupy some sites in common,
these two loops are regarded as one loop with the number of crossings.
It is found that 
in the confinement phase, a long monopole loop exists in each configuration 
in addition to some short loops.
The separation of long loop and other short loops is 
clearly seen in Table \ref{t1}, where
the data are obtained from 10 configurations at $\beta=$ 2.20 
on $16^3 \times 4$ lattice. Note that
the number of long loop is equal to the number of configurations.
In this case,  $\beta_c$ is 2.298. 
For $\beta <\beta_c$  long monopole loops have a characteristic length
at each $\beta$ and the long loops become shorter or split into a few parts 
around $\beta_c$.
In the deep deconfinement region, no long loop exists and 
all monopole loops are short.
These results are consistent with those in the $U(1)$ case.\cite{bode}

%---------- table 2 --------------
\begin{table}[tbh]
\begin{center}
\begin{tabular}{c|rrrrrrrr}
(a)                     &       &      &      &      &      &     
                        &       & \\       
$\beta$                 &  2.35 & 2.45 & 2.48 & 2.51 & 2.53 & 2.57 
                        &  2.63 & 2.70 \\ \hline
$\langle L\rangle$      &  4980 & 2205 & 1710 &  838 &  479 & 148  
                        &   69  & 35   \\ \hline
$\langle$crossing$\rangle$ &  1120 & 316 & 219 &  94 &   50 &  13  
                        &  5    &  2  \\ \hline
charge 1       & 97.62     & 98.67 & 98.88 & 99.04 & 99.07 & 99.52 
               & 99.29     & 99.27    \\ 
charge 2       &  2.36     &  1.32 &  1.12 &  0.96 &  0.93 &  0.48 
               &  0.70     &  0.73    \\ 
charge 3       &  0.02     &  0.01 &  0.00 &  0.00 &  0.00 &  0.00 
               &  0.00     &  0.00    \\ 
charge 4    &  0.00     &  0.00 &  0.00 &  0.00 &  0.00 &  0.00 
               &  0.00  &  0.00       \\ \hline\hline
(b)                     &       &      &      &      &      &      
                        &       &      \\       
$\beta$                 &  2.35 & 2.45 & 2.48 & 2.51 & 2.53 & 2.57 
                        &  2.63 & 2.70 \\ \hline
$\langle L\rangle$      &  1140 &  700 &  605 &  450 &  357 & 169  
                        &   54  & 21   \\ \hline
$\langle$crossing$\rangle$& 714 &  320 &  246 &  157 &  110 &  45  
                          & 12  &  3   \\ \hline
charge 1       & 81.07     & 90.98 & 92.10 & 93.85 & 94.92 & 95.92 
                           & 96.44 & 98.54 \\ 
charge 2       & 16.98     &  8.43 &  7.69 &  5.91 &  5.02 &  4.02 
                           &  3.36 & 1.45  \\ 
charge 3       &  1.87     &  0.57 &  0.20 &  0.23 &  0.06 &  0.06 
                           &  0.19 &  0.00 \\ 
charge 4       &  0.09     &  0.03 &  0.01 &  0.01 &  0.00 &  0.00 
                           &  0.00 &  0.00 \\ 
\end{tabular}
\end{center}
\caption{
The length of the longest loop, the number of crossings 
and the ratio (percent) of the number of multicharged monopole currents 
to the total number of monopole currents in the case of 
$2^3$-extended monopoles (a) and that of $4^3$-extended monopoles (b) 
on $24^3\times 8$ lattice.
The numbers are averages over 30 configurations.
}
\label{t2}
\end{table}
%---------------------------------

In Table \ref{t2}, the $\beta$-dependence of the loop length 
$\langle L\rangle$, 
the number of crossings and the ratio of the number of multicharged 
currents to the number of the total currents are shown.
Those quantities are calculated from 30 
configurations for two different sized monopoles on $24^3\times 8$ lattice.
$\beta_c$ is 2.51 on this lattice.
The number of crossings is small in the case of $2^3$ monopole, 
whereas it is not so small in  the case of $4^3$ monopole.
The monopole currents with a unit charge are dominant for both extended 
monopoles, which is consistent with 
that $f_1$ is dominant on the monopole action.
In the case of $4^3$ monopole, $\langle L\rangle$ exceeds the site number of 
the renormarized lattice for $\beta<\beta_c$, 
whereas they are nearly equal around $\beta_c$.
In the case of $2^3$ monopole $\langle L\rangle$ is smaller than 
the number of the sites.

Recently, Ejiri et al.\cite{ejiri2,ejiri3} have evaluated the string 
tension derived from long and short monopole loops separately.
They have found that the long loops alone can reproduce 
the full value of the string tension in spite of the fact that 
only a few percents of a total dual links are occupied by the long loop.
Other short loops do not contribute 
 to the string tension.
This suggests that the long loops are essential 
in the deconfinement transition.

\subsection{The distribution of monopole currents}
Now we pay attention to long loops and 
investigate the distribution of monopole currents contained 
in a long loop.
Even the longest loop occupies only about 10 percent of the total 
dual links.
Does the long loop spread uniformly through the lattice or not?
In Ref.\citenum{bode}), it is filling up the whole lattice in compact QED.

%----------- Fig.9----------------
\begin{figure}
\vspace{-1cm}
\epsfxsize=0.6\textwidth
\begin{center}
\leavevmode
\epsfbox{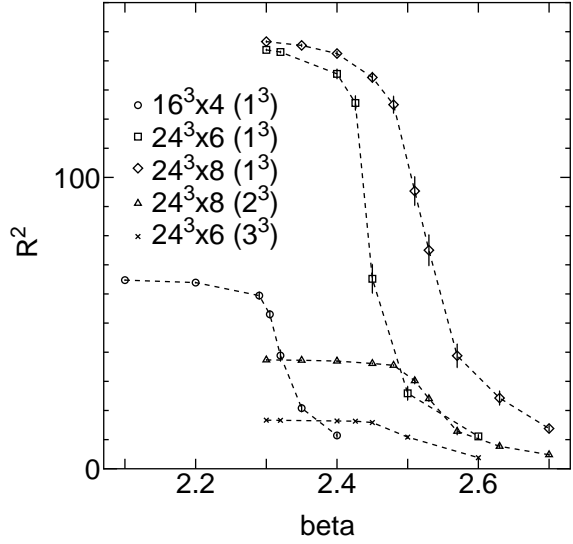}
\end{center}
\vspace{-1cm}
\caption{
$R^2$ versus $\beta$ on various lattices.
The number in the parentheses represents the extendedness of monopoles.
}
\label{f9}
\end{figure}
%---------------------------------

We calculate the mean square of the distance from the center as follows: 
\begin{eqnarray}
    R^2=\frac{1}{L}\sum_{i} (\vec{r}_i-\vec{R}_G)^2 ,
\end{eqnarray}
where $L$ is the length of the monopole loop, $\vec{r}_i$ is the 
position of a monopole current and $\vec{R}_G$ is the position 
of the center.
If the monopole currents exist uniformly on the $N_s^3\times N_t$ lattice, 
$R^2$ should be evaluated as
\begin{eqnarray}
   \frac{1}{N_s^3N_t}\int^{N_s/2}_{-N_s/2}dx\int^{N_s/2}_{-N_s/2}dy
      \int^{N_s/2}_{-N_s/2}dz\int^{N_t/2}_{-N_t/2}dt
      (x^2+y^2+z^2+t^2) 
   =\frac{N_s^2}{4}+\frac{N_t^2}{12}   .
\end{eqnarray}
In the case of $n^3$ extended monopole, $R^2$ should be
\begin{eqnarray}
   \frac{(N_s/n)^2}{4}+\frac{(N_t/n)^2}{12}. \label{r2}
\end{eqnarray}
Our data of $R^2$ are shown in Fig.\ref{f9}.
$\beta_c$ is 2.426 in the case of $N_t=6$ 
and $\beta_c$ is 2.51 in the case of $N_t=8$. 
In the confinement phase, the data almost coincide with the expected 
values (\ref{r2}).
For example, $R^2$ is expected to be 147 
 in the case of $1^3$ monopole on $24^3\times 6$ lattice
and in the case of $3^3$ monopole 
$R^2$ is expected to be 16. 
The long monopole loop is almost uniformly distributed 
in the whole lattice in the confinement phase.
In the deconfinement phase $R^2$ starts to decrease rapidly 
above $\beta_c$.
In the case of extended monopoles the uniformity is better 
than in the $1^3$ case for $\beta \leq \beta_c$.

\subsection{Monopole effective size}
Assuming that crossings of the loop are ignored, 
 an $n^3$ monopole loop seems to spread 
almost uniformly 
through the whole lattice for $\beta\leq\beta_c$.
Considering that almost all magnetic charge is $\pm 1$ and that 
the loop length is rather restricted in comparison with  all link 
numbers $4\times N_s^3\times N_t$, there appears to work 
some repulsive force between currents. To study the force,
we may define an effective size of monopoles as follows:
\begin{eqnarray}
   l^3(n)=\frac{(N_s/n)^3\times (N_t/n)}
               {\langle L(n)\rangle} ,
\end{eqnarray}
where the numerator is the volume of the renormalized lattice 
of $n^3$ extended monopoles 
and the denominator is the length of the monopole loop 
in the renormalized lattice unit.
For $\beta > \beta_c$ the distribution of monopole currents 
is not uniform as seen in the previous subsection.
In this case we  assume 
that the distribution is uniform in the region 
restricted to $2R$ in each spatial direction and $N_t$ 
in the time direction. Note that 
 $2R$ is still larger than $N_t$ near $\beta_c$. Hence 
the effective renormalized volume which is occupied by the longest loop 
may be given by $(2R/n)^3 \times (N_t/n)$.
$l(n)$ represents the range affected by one of the $n^3$ monopole 
world line.
If the monopole current with a range $l(n)$ exists on a link, 
the nearest current could be put on links  more apart 
than $l(n)$.
The $n^3$ extended monopole current with a range $l(n)$ seems to have 
an exclusion volume $l(n)^4$ in the renormalized lattice.

Considering the exclusion volume,
the entropy of monopoles may be discussed, 
in the case of $n^3$ extended monopoles having the range $l(n)$, 
not on the whole renormalized lattice but
on a reduced lattice 
\begin{eqnarray}
   \frac{(N_s/n)^3\times (N_t/n)}{N(l)^4} 
\quad ({\rm for } \beta\le\beta_c),
\end{eqnarray}
where $N(l)$ is an integer depending on $l(n)$. 
It seems natural to adopt $N(l)= [l(n)]+1$, 
where the symbol $[l(n)]$
is the integer not exceeding $l(n)$.
Monopoles are seen 
as if they were running  only on the reduced lattice.

%----------- Fig.10----------------
\begin{figure}
\vspace{-1cm}
\epsfxsize=0.6\textwidth
\begin{center}
\leavevmode
\epsfbox{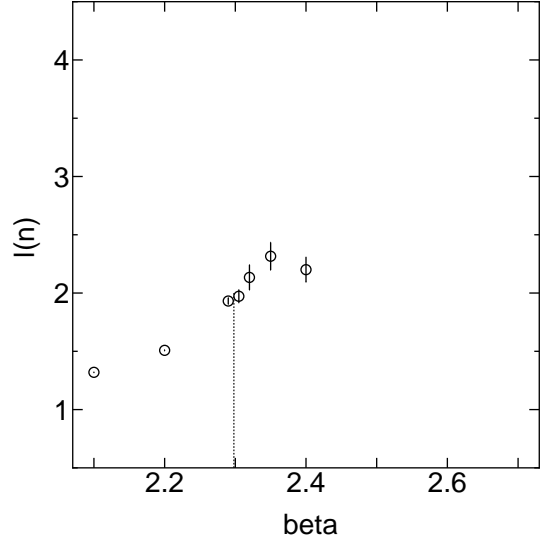}
\end{center}
\vspace{-1cm}
\caption{
$l(n)$ versus $\beta$ for $1^3$ monopoles on $16^3\times 4$ lattice.
The dotted vertical line denotes $\beta_c$.
}
\label{f10}
\end{figure}
%---------------------------------

%----------- Fig.11----------------
\begin{figure}
\epsfxsize=0.6\textwidth
\begin{center}
\leavevmode
\epsfbox{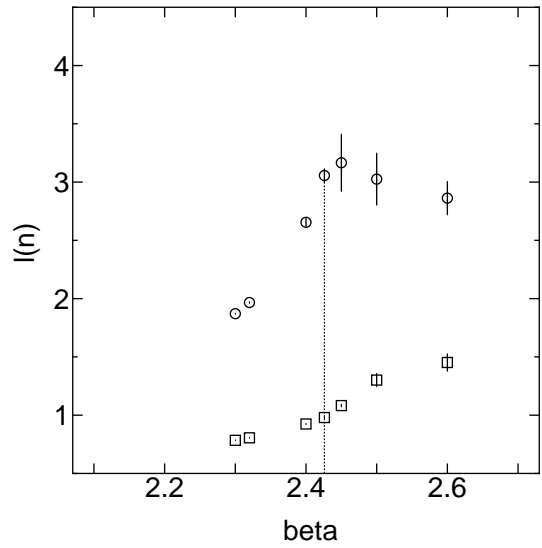}
\end{center}
\vspace{-1cm}
\caption{
$l(n)$ versus $\beta$ for $1^3$ (circle) and $3^3$ (square) monopoles 
on $24^3\times 6$ lattice.
The dotted vertical line denotes $\beta_c$.
}
\label{f11}
\end{figure}
%---------------------------------

%----------- Fig.12----------------
\begin{figure}
\vspace{-1cm}
\epsfxsize=0.6\textwidth
\begin{center}
\leavevmode
\epsfbox{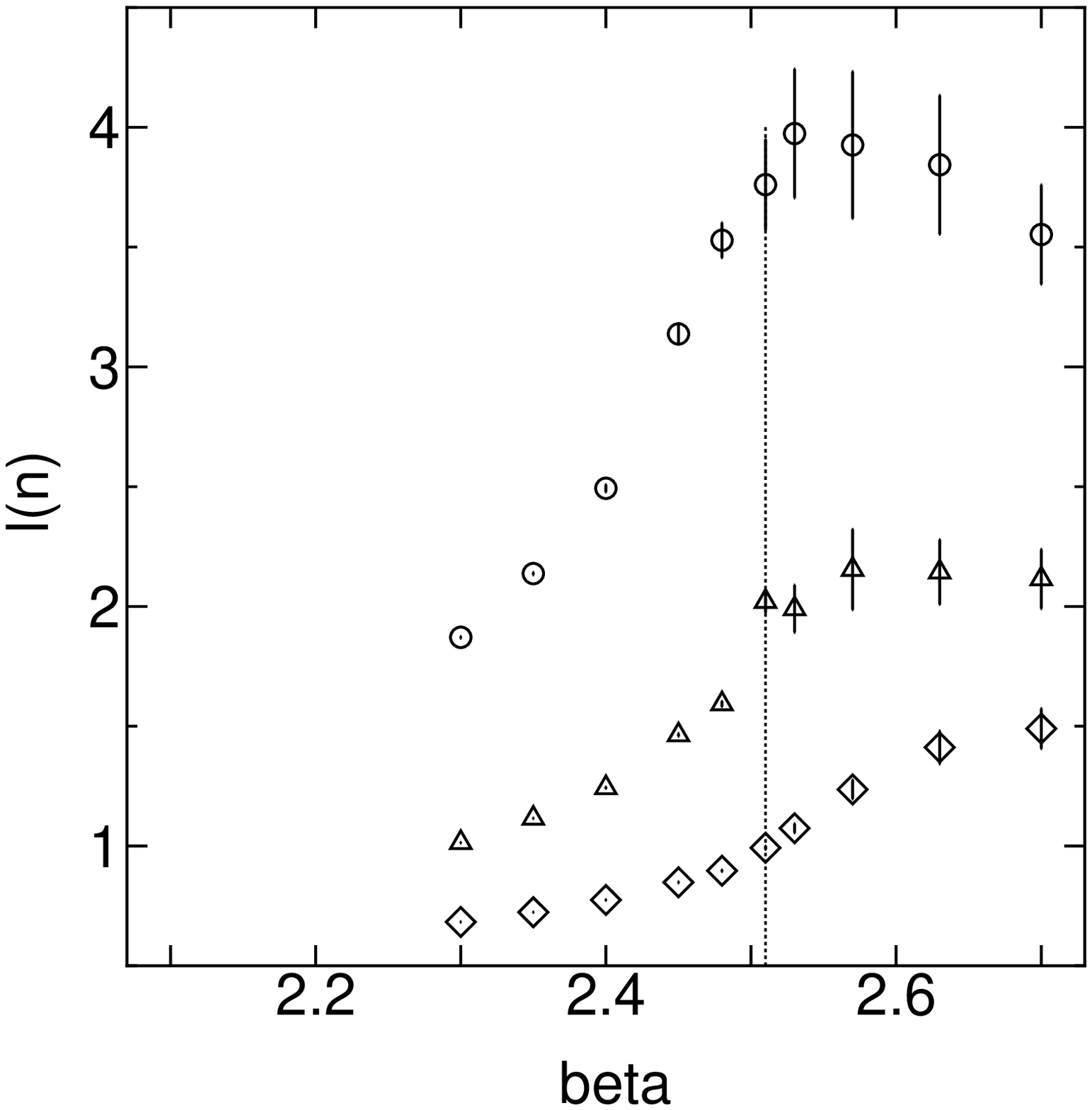}
\end{center}
\vspace{-1cm}
\caption{
$l(n)$ versus $\beta$ for $1^3$ (circle), $2^3$ (triangle) and $4^3$ (diamond) 
monopoles on $24^3\times 8$ lattice.
The dotted vertical line denotes $\beta_c$.
}
\label{f12}
\end{figure}
%---------------------------------

$l(n)$ are shown in Figs.\ref{f10}-\ref{f12}.
In the case of $1^3$ monopole on $16^3\times 4$ lattice shown 
in Fig.\ref{f10}, 
we see that $l(n=1)$ is less than 2 for 
$\beta <\beta_c$ and approaches 2 at $\beta_c$.
$l(n=1)$ approaches 3 from below as $\beta \to \beta_c$ 
in the case of $1^3$ monopole 
 on $24^3\times 6$ lattice, whereas  $l(n=3)\le 1$ in the case of $3^3$ 
 as seen from Fig.\ref{f11}.
The same regularity exists also 
for any extended monopole on $24^3\times 8$ 
lattice shown in Fig.\ref{f12}.
It is stressed that
$l(n)$ is always less than or equal to $N_t/(2n)$ for $\beta\le\beta_c$.
This leads us to 
 that the reduced lattice always has a
time extent $2= (N_t/n)/(N_t/(2n))$  for $\beta\le\beta_c$.
Then the entropy of a $n^3$ extended monopole loop for any $n$ 
should be calculated always on a reduced lattice with the time extent 2
for $\beta\le\beta_c$.
This means that 
{\it
the entropy of a $n^3$ monopole is the same for any $n$, since the 
reduced lattice is the same for any extendedness $n$ 
if the original lattice is the same.
}

\subsection{The entropy of a monopole loop and the deconfinement transition}
In the previous subsection, we have seen that
the entropy of a $n^3$ monopole is the same for any $n$ 
if the original lattice is the same. However, the action is different 
for different extended monopoles at  fixed $\beta$. Since the entropy is 
equal, larger extended monopoles are more important to fix the 
point of the deconfinement transition as seen from Fig.\ref{f7}. 

What is the maximum size of extended monopole?
To define dynamical monopole currents which are important,
 the renormalized lattice should have 
at least two independent lattice sites in the time direction.
This means that the maximum size $n_{max}$ of dynamical extended monopoles 
is $N_t/2$. From Figs.\ref{f10}-\ref{f12}, we see the maximally extended 
monopoles have always $l(n_{max})\le 1$ 
for $\beta\le\beta_c$. 
The reduced  lattice is the same as the renormalized one 
for $\beta\le\beta_c$ 
in the case of the maximally extended monopole.

What is the entropy of a long monopole loop on the reduced lattice 
with $N_t=2$ ? As shown above,
a long monopole loop seems to behave 
like a random walk with a stronger repulsive condition than a simple 
non-backtracking.
However to know the condition correctly is not simple.
Moreover, the number of crossing increases 
as extendedness becomes large. 
The entropy is very difficult to estimate contrary to the case
 in $T=0$ $SU(2)$ QCD.

%----------- Fig.13----------------
\begin{figure}[tbh]
\vspace{-1cm}
\epsfxsize=0.8\textwidth
\begin{center}
\leavevmode
\epsfbox{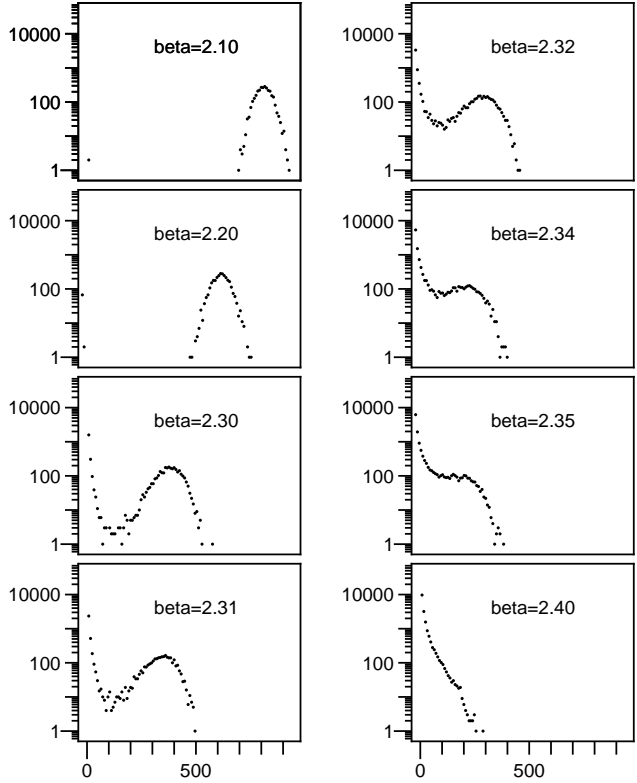}
\end{center}
\vspace{-1cm}
\caption{
The histograms of the length of $2^3$ monopole loops 
on $12^3\times 4$ lattices.
}
\label{f13}
\end{figure}
%---------------------------------

Therefore, let us
investigate the entropy by a histogram method using Monte Carlo simulations.
It is easy to get the histogram of the length of monopole loops from
many configurations generated in the Monte Carlo simulations.
The lengths of $2^3$ extended monopole loops are measured on 
$12^3\times 4$ lattice, since they are the maximally extended monopole in this 
case.
The histogram is obtained from 3000 configurations as shown in Fig.\ref{f13}.
$\beta_c$ is 2.298 on the lattice.
The vertical line represent the frequency of length.
The distribution of the long monopole loops seems to be Gaussian. 

The number of loops of length $L$ through a given point goes 
asymptotically like $\mu^L$ irrespective of the conditions of 
random walks.\cite{gans} Hence 
 the distribution of the long loop length may be fitted by the following 
form 
\begin{eqnarray}
   -aL^2+ bL+ c,  \quad b= \ln\mu -f_1 ,
\end{eqnarray}
where $f_1$ comes from the self energy term 
in the monopole action and c is a constant.
The term of $-aL^2$ should come from 
all other terms than the self energy and it is  very small as seen 
from Fig.\ref{f8}. But the term is important 
because it suppresses proliferation 
of monopole loops in the condensed phase. 
The linear term may determine the entropy $\ln \mu$.

%-------------- table 3 ----------------
\begin{table}[tbh]
\begin{center}
\begin{tabular}{ccc}
 $\beta$ & a                    & b                    \\ \hline
 $2.10$  & $1.80\times 10^{-4}$ & $2.93\times 10^{-1}$ \\
 $2.20$  & $1.41\times 10^{-4}$ & $1.76\times 10^{-1}$ \\
 $2.30$  & $6.65\times 10^{-5}$ & $4.93\times 10^{-2}$ \\
 $2.35$  & $5.70\times 10^{-5}$ & $2.31\times 10^{-2}$ \\
\end{tabular}
\end{center}
\caption{
Least square fitting of the histogram for $\beta=2.10$, $2.20$, $2.30$ 
and $2.35$.
}
\label{t3}
\end{table}
%-------------------------------------

We plot in Table \ref{t3} 
the values of the parameters of the Gaussian fit to the data at 
$\beta =2.1 \sim 2.35$, where the Gaussian shape is seen.
We see from the table that the linear term at $\beta_c= 2.298$ is already
small and it almost vanishes 
around $\beta=$ 2.35.
If the balancing of energy and entropy of maximally extended monopole 
is the origin of the deconfinement transition, 
the predicted critical coupling is $\beta_c \sim 2.35$ 
which is near to the actual critical 
coupling $\beta_c= 2.298$. 
Considering the finite size effects near $\beta_c$, the agreement is 
fairly good.

From the fit, one can estimate roughly 
the value of the entropy using the values 
of $f_1 (\beta)$, i.e., $f_1 (2.35)\sim 1.3$ and $f_1 (2.298)\sim 1.16$. 
We get $1.16 \le \ln \mu \le 1.3$.

%----------- Fig.14----------------
\begin{figure}[tbh]
%\vspace{-3cm}
\epsfxsize=0.6\textwidth
\begin{center}
\leavevmode
\epsfbox{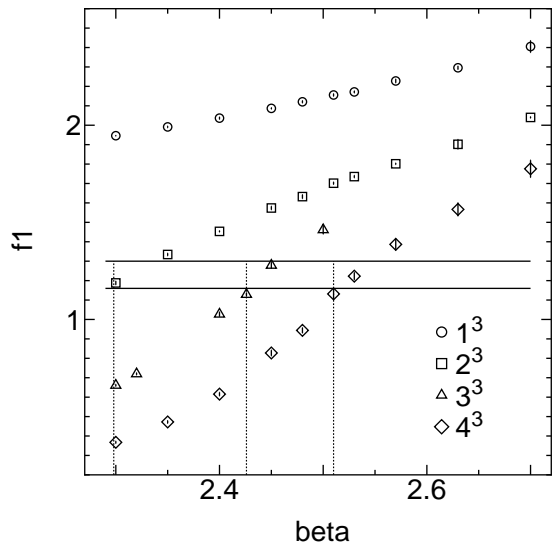}
\end{center}
\vspace{-1cm}
\caption{
Coupling constant $f_1$ for various extended monopoles.
The dotted vertical lines denote $\beta_c$ for $N_t =4, 6$ and $8$ from the left.
}
\label{f14}
\end{figure}
%---------------------------------

In Fig.\ref{f14}, we plot $f_1$ for maximally extended monopoles on 
various lattices in comparison with the entropy estimated above.
We find that the balancing of the 
energy and the entropy of the monopoles 
can explain the transition points of all lattices approximately.
There are, however, many things to be studied. 
Nevertheless, 
 it is very interesting that 
{\it the deconfinement transition can be 
understood roughly by the simple idea of balancing of energy and entropy 
of monopole loops in $T\neq 0$ $SU(2)$ QCD.
}

%\begin{center}
\section{Conclusion and remarks}
%\end{center}
\begin{enumerate}
\item
In $T\neq 0$ QCD, it is found that a long monopole loop exists 
in each configuration and all other loops are short in the confinement phase.
They distribute almost uniformly in the confinement phase.
Table \ref{t1} shows that 
the number of short loops decrease in the case of $2^3$ monopoles 
in comparison with the case of $1^3$ monopoles.
To study the continuum limit, we have to make the time extent $N_t$ 
large. Then large extended monopoles are important and 
short loops would disappear rapidly. It is expected that 
long monopole loops alone are responsible for the confinement mechanism 
 of quarks. This is consistent with the result that
the string tension is reproduced only by long monopole 
loops.\cite{ejiri2,ejiri3}
\item
Almost all magnetic charge is $\pm 1$ and  
the loop length is rather restricted in comparison with  all link 
numbers of lattice.
There appears to work 
some repulsive force between currents, so that monopoles may have 
an exclusion volume around them. Considering the exclusion volume
determined from the distribution data,
we have derived that the entropy of extended monopoles may be independent of 
the extendedness if the original lattice is the same.
\item
In $T\neq 0$ QCD there exists a maximally extended 
($(N_t/2)^3$) monopole contrary to the case of 
$T=0$ $SU(2)$ QCD where   
 larger and larger extended monopoles become important. 
The critical temperature becomes necessarily finite. 
Balancing of  energy and entropy 
of the maximally extended monopoles can explain roughly 
the deconfinement transition of various lattices. 
However more intensive studies are 
needed to clarify the transition mechanism definitely.
Especially, the finite size effect near $\beta_c$ should be 
clarified in details.

\end{enumerate}

\vspace{2cm}

%\begin{center}
\acknowledgements
%\end{center}
The authors are thankful to Hiroshi Shiba and  Shinji Ejiri for 
fruitful discussions.
This work is financially supported by JSPS Grant-in Aid for 
Scientific  Research (B)\\ 
(No.06452028).

\end{document}